\documentclass[12pt]{iopart}
\usepackage[pdftex]{graphicx}
\usepackage{dcolumn}
\usepackage{bm}
\usepackage{array}
\usepackage{color}
\usepackage{float}
\usepackage{subfigure}
\usepackage{dsfont}
\usepackage{txfonts}
\usepackage{wasysym}
\usepackage{multirow}
\usepackage{sidecap}
\usepackage{xcolor,cancel}
\usepackage[normalem]{ulem}
\usepackage{cite}

\newcommand{\dlangle}{\langle\langle}

\newcommand{\+}{\dagger}

\newcommand{\dn}{\downarrow}
\newcommand{\drangle}{\rangle\rangle}

\newcommand{\ve}{\varepsilon}

\newcommand{\SC}{\mathrm{SC}}
\newcommand{\leads}{\mathrm{leads}}

\newcommand{\qdot}{\mathrm{dots}}
\newcommand{\dotleads}{\mathrm{dot-leads}}
\newcommand{\dotSC}{\mathrm{dot-SC}}
\newcommand{\T}{\mathrm{T}}

\newcommand{\up}{\uparrow}

\newcommand{\veck}{\mathbf{k}}

\newcommand{\meV}{{\rm meV}}

\begin{document}
%
\title{Andreev and Majorana bound states in single and  double quantum dot 
structures}

\author{Joelson F. Silva and E. Vernek}

\address{Instituto de F\'isica, Universidade Federal de Uberl\^andia, 
Uberl\^andia, Minas Gerais 38400-902, Brazil.}
\ead{vernek@ufu.br}

\date{\today}
\begin{abstract}
We present a numerical study of the emergence of Majorana and Andreev 
bound states in a system composed by two quantum dots, in which one of then is 
coupled to a conventional superconductor, SC1, and the other connects to a 
topological superconductor, SC2. By controlling the interdot coupling we can 
drive the system from  a two single (uncoupled) quantum dots to a double 
(coupled) dot system configurations. We employ a recursive Green's 
function technique that provides us with numerically exact results for the  
local density of states  of the system. We first show that in the uncoupled 
dot configuration (single dot behavior) the Majorana and the Andreev bound 
states appear in an individual dot in two completely distinct regimes. 
Therefore, they cannot coexist in the single quantum dot system.  We 
then study the coexistence of these states in the coupled double-dot 
configuration. In this situation we show that in the trivial phase of the SC2, 
the Andreev states are bound to an individual quantum dot in the atomic regime 
(weak interdot coupling)  or extended over the entire molecule in the molecular 
regime (strong interdot coupling). More interesting features is actually seen in 
the topological phase of the SC2. In this case, in the atomic limit, the 
Andreev states appear bound to  one of the quantum dots while a Majorana zero 
mode appears in the other one. In the molecular regime,  on the other hand, the 
Andreev bound states take over the entire molecule while the Majorana state 
remains always bound to one of the quantum dots.

\end{abstract}
\date{\today}
\maketitle
\ioptwocol
\section{Introduction}
Nanoscopic structures coupled to superconductors has attracted the 
attention of many researchers over the last two 
decades\cite{Vieira1,Vieira2,Buitelaar,Nazarov,Maurand,Trocha,Goffman,
Wernsdorfer,Pascual}. One interesting aspect of nanostructured systems such as 
quantum dots coupled to normal or superconducting contacts is their flexibility 
to study many different phenomena, such as the Kondo 
effect\cite{Goldhaber-Gordon}, the Andreev bound states (ABS), etc. Due to 
confinement,  quantum dot systems exhibit a discrete spectrum  that can be 
modified by local gates, allowing for a fine control of their physical 
properties. 

When a quantum dot is connected to a conventional superconductor, 
multiple Andreev reflections between the quantum dot and the superconductor 
give rise to Andreev bound states (ABS) in the dot\cite{Balatsky,Baranski}, 
 with signature visible in their transport properties\cite{Kim,Dirks}. 
For example, by studying a graphene quantum dot coupled to 
a normal and to a superconductor contacts,  Dirks  et al\cite{Dirks} have 
observed sharp {\it in gap} conductance peaks attributed to individual ABS in 
the system.  Another interesting result is the interplay between the Kondo 
screening and the superconducting pairing in quantum dots observed by Pillet et 
al\cite{Goffman}. Interplay between Kondo and Andreev bound states has 
also been studied earlier by  Franke and coworkers\cite{Pascual}, in magnetic 
molecules on the surface of a superconductor. 
 
More recently,  quantum dots coupled to topological 
superconductor wires\cite{Golub,Liu,Lee,Liu2}  supporting Majorana bound states 
(MBS) in their ends have also attracted a lot of attention. Besides the 
interest in the fundamental concepts of the Majorana state 
physics\cite{Read,Alicea,Kitaev}, it is also potentially useful for 
topologically protected quantum 
computation\cite{Pachos,Sarma1,Akhmerov,Laflamme,Oreg1,Nayak}. Despite  the 
great effort of various researchers toward experimental 
observation\cite{Mourik,Deng,Das}, no much progress have been made 
lately. One of the most challenging task for a clear 
observation of MBS features in condensed matter systems is to tell apart 
features from  other phenomena, as it has been argued that MBS may be confused 
with signatures from Kondo or ABS. For instance, it has been 
shown by one of us that, under certain regime, the Majorana and the Kondo 
physics can coexist in a quantum dot that is simultaneously coupled to a 
topological superconductor and to a metallic contact\cite{Lee,Vernek1}. 
Therefore, identifying features of a particular phenomena can be 
experimentally puzzling. While the interplay between the Majorana and the 
Kondo effect have been investigated in some detail, very little attention have 
been devoted to  MBS and ABS together\cite{Chevallier,Golub1}. For 
instance, do MBS and ABS mix together? More importantly, if the MBS and ABS 
coexist in some regime, how can one distinguish them experimentally? The 
answer to these important questions still remain unclear. 
\begin{figure}[b!]
\centering
\subfigure{\includegraphics[clip,width=3.4in]{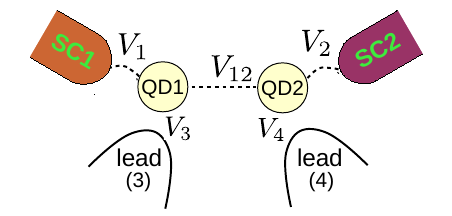}}
\caption{(Color online) Schematic representation of the system. 
SC1 and SC2 represent a conventional $s$-wave  and a topological 
superconductor, respectively. The terminals 3 and 4 are normal metallic leads 
that introduces a natural broadening of each quantum dot spectrum. QD1 and 
QD2 represents single level and non-interacting quantum dots. By 
controlling $V_{12}$ we can drive the system from a two single (uncoupled) dots 
to a double (coupled) dot configurations.} 
\label{model}
\end{figure} 

Aiming to addressing these questions, in this work we investigate possible 
coexistence and interplay between Andreev and Majorana bound states in 
a single and double dot systems. We consider a double quantum dot system in 
which one of the dots is  coupled to a conventional superconductor while the 
other couples to topological superconductor. The quantum dot are interconnected 
by a controllable tunneling barrier that allows us to drive the system from a 
single to  a double dot configurations,  providing a flexible structure to 
address  possible mixing of MBS and ABS. For the sake of clarity, our system, 
schematically represented in Fig.~\ref{model}. It resembles the one proposed by 
Pillet et al, in Ref.~\cite{Goffman}. Here, however, while the left quantum dot 
(QD1) is coupled to a normal $s$-wave superconductor (SC1) the right one (QD2) 
is connected to a topological superconductor that provides the Majorana bound 
state. Furthermore, each of the QDs are coupled to individual metallic leads 
labeled by (3) and (4) (see Fig.~\ref{model}) that induces a natural 
broadening of the QD levels. When the QDs are  decoupled from each 
other, if  the SC2 is in its topological phase, a Majorana mode will appear in 
the QD2, as predicted in Ref.~\cite{Vernek1}, while in the QD1 there will be 
a pair o ABSs. 

To study the bound states we focus mainly on the spectral properties of 
the system, that is accessible via zero-bias differential conductance 
measurement as in experiment reported in Ref.~\cite{Goffman}. 
We employ the  standard  recursive Green's function approach, from 
which we can obtain numerically exact results for the local density of states 
of the system. We first study the situation in which the QDs are fully 
decoupled from each other. In this configuration  we monitor the local 
density of states of the QD2 while driving the SC2 from its trivial to its 
topological phase. We show that the MBS and the  ABSs appear in two distinct 
regimes, that can be distinguished in tunneling  spectroscopy measurements. 
We then go on and study possible coexistence of MBS and ABS in the double dot 
configuration. To do so we couple the two QDs by making the interdot coupling 
finite. We find  that in the trivial phase of the SC2 and in the atomic limit 
(small interdot coupling) the dot coupled to the normal superconductor exhibits 
the usual Andreev bound states whereas the other one exhibits a atomic single 
particle peak. In the molecular regime (strong interdot coupling), the Andreev 
states take over the entire molecule, appearing clearly in the molecular 
orbitals. More interestingly, in the topological phase of the SC2, in the atomic 
regime we see Andreev bound states in the QD1 and a Majorana zero mode in the 
QD2. In the molecular regime, on the other hand, while the Andreev bound states 
is seen over entire molecule, the  Majorana mode (seen only for spin 
\emph{down}) is always bound to an atomic orbital.  


The remainder  of this paper is structured in the following 
way. In Sec.~\ref{sec:model} we describe the model and derive the main physical 
quantities and in Sec.~\ref{Numerical} we present our numerical results. 
Finally, we conclude our work in Sec.~\ref{Conclusion}. 

\section{Model and method}\label{sec:model}
\subsection{Hamiltonian model}
For concreteness, our system is described by the following Hamiltonian:
\begin{eqnarray}\label{Ful_H}
 H&=&H_{\qdot}+H_{\SC}+H_{\leads}+H_{\dotleads} +H_{\dotSC}\nonumber\\
 &&+H_{\T}.
\end{eqnarray}
Here $H_{\qdot}$ and  $H_\leads$ describes the isolated quantum 
dots and the normal leads, respectively, $H_\SC=H_1+H_2$ describes the normal 
and the topological superconductors, $H_\dotleads$ connects the dot to the 
normal leads, $H_\dotSC$ couples the dots the superconductors and $H_{\T}$ 
describes the tunnel coupling between the dots.  Explicitly, the various 
terms in the   Hamiltonian (\ref{Ful_H}) can be written as, 
\begin{eqnarray}\label{H_dot}
 H_{\qdot}=\sum_{i=1,2\atop s}\ve_{i,s}d^\+_{i, s}d_{i, s},
\end{eqnarray}
\begin{eqnarray}
 H_1=\sum_{j=-\infty\atop s}^{-1}\left(t_1 c^\dagger_{j-1,s}c_{j,s}+\Delta_1 
c_{j,\up}^\+c^\+_{j,\dn} +{\rm H.c.}\right),
\end{eqnarray}

\begin{eqnarray}\label{H2}
H_2&\!=\!&\frac{1}{2}\sum_{j=1, s}^\infty\left[\left(-\mu+V_Z\sigma^z_{ s s}  
\right)c^\dagger_ { j, s }c_{j, s}-t_2c^\dagger_{j+1, s}c_{j, s } 
\right.\nonumber\\
&&\left. +\Delta_2 c^\+_{j,\up}c^\+_{j,\dn}\right]+it_{\rm 
SO}\sum_{j=1\atop s s^\prime}^{\infty}(c^\+_{j+1, s}
\sigma^{y}_{ss^\prime}c_{j, s^\prime})+{\rm H.c.},\nonumber \\
\end{eqnarray}
\begin{eqnarray}
H_{\leads}=\sum_{\ell \veck,s}\ve_{\ell \veck, s}c^\+_{\ell \veck, s}c_{\ell 
\veck, s} \quad \mbox{($\ell=3,4$)},
\end{eqnarray}
\begin{eqnarray}
H_\dotleads\!=\!\sum_{\veck, s}\left(V_{3}d^\+_{1, s}c_{3, 
\veck, s}+V_{4}d^\+_{2, s}c_{4,
\veck, s}\!+\! {\rm H.c.}\!\right),
\end{eqnarray}
%
%
\begin{eqnarray}
H_{\dotSC}\!=\!-\sum_{s}\left(V_1 d^\+_{1,s}c_{-1,s}+ V_2 
d^\+_{2,s}c_{1,s}\right)+{\rm H.c.},
\end{eqnarray}
and 
\begin{eqnarray}\label{H_T}
H_{\T}=-V_{12}\sum_{s}\left(d^\+_{1, s}d_{2, s}+d^\+_{2, s}d_{1, 
s}\right).
\end{eqnarray}
In the Eqs.~(\ref{H_dot})-(\ref{H_T}),  $d^\+_{i,s}$ ($d_{i,s}$) creates 
(annihilates) an electron  with energy $\ve_{i,s}$ and spin $s$ in the 
quantum dot $i=1,2$, $c^\+_{j,s}$ ($c_{j,s}$) creates (annihilates) an 
electron with spin $s$ in the $j$-th site of the superconductors and 
$c^\+_{\ell \veck}$ ($c_{\ell \veck}$) creates (annihilates) an electron 
with energy $\ve_{\ell,\veck}$ with spin $s$ in the $\ell$-th lead. Note 
that, besides the $s$-wave pairing potential $\Delta_2$ of $H_2$, it also 
contains a Rashba spin-orbit interaction characterized by the $t_{\rm SO}$ and 
 the effect of an external magnetic field that produces the Zeeman energy 
splitting $V_z$. These three ingredients are important for the topological 
regime of the superconductor\cite{Roman,Mourik}. The coupling $V_{3(4)}$ are 
assumed to be $\veck$-independent, for simplicity, and the leads are considered 
to be identical and characterized by a flat density of states 
$\rho_0(\omega)=(1/2D)\Theta(D-|\omega|)$, where $D$ is their half bandwidth. 
In the wide band limit, the influence of the lead 
on the dots are just a broadening $\Gamma=\pi V^2/2D$ in the QD levels, where 
$V_3=V_4=V$. The wide band limit for the leads assumed 
here just simplifies the equations but it is not mandatory. In 
this limit, our calculation fully accounts the effect of the leads.

The topological phase of the SC2, described by the Hamiltonian 
(\ref{H2}), is obtained\cite{Alicea2} for $V_Z>V_Z^{c}$, where 
$V_Z^c=\sqrt{\tilde\mu^2+\Delta_2^2}$, with $\tilde\mu=\mu+t$.   In 
this phase, the SC2 holds one Majorana mode at each of its ends. In our case, 
one of the ends of the topological wire is coupled to QD2 and the Majorana mode 
leaks into it, as we have shown in recent studies\cite{Vernek1,Vernek2}. On 
the other hand, for $V_Z<V_Z^c$ the wire is in its trivial phase and no 
Majorana will be present.

{\subsection{Electron Green's function}}\label{Electron-GF}
To study the local density of states (LDOS) of the QDs we employ the Green's 
function (GF) method. Using the equation of motion techniques we obtain a 
recursive expression for the GFs. We follow a similar approach as described in 
Ref.~\cite{Vernek2} and define the retarded GF matrix as
\begin{eqnarray}\label{GFmatrix}
{\bf G}_{i,j}(\varepsilon) \equiv \dlangle A_i;B_j \drangle,
\end{eqnarray}
where $A_i$ and $B_i$ (with $i$ and $j$ denoting any site of the system 
(including the quantum dots) are any of the operators $c_{i\uparrow}, 
c_{i\downarrow},c^\dagger_{i\uparrow},c^\dagger_{i\downarrow}$ 
and the double bracket stands for the Fourier transform of the double time 
GF defined  as\cite{Zubarev} 
\begin{eqnarray}
 \dlangle A_i;B_j \drangle \!=\!\!\int_{-\infty}^\infty \!\!
(-i)\Theta(\tau)\langle [A_i(t),B_j(t^\prime)]_+\rangle 
e^{i\varepsilon \tau} d \tau.
\end{eqnarray}
Here $\tau=t-t^\prime$ and $[\cdot,\cdot]_+$ denotes the anticommutator 
between two fermion operators. The GF (\ref{GFmatrix})  defines a $4\times 
4$ matrix  that is determined via energy space equation of motion for its 
elements
\begin{eqnarray}\label{EOM}
 \varepsilon \dlangle A_i,B_j \drangle=\langle [A_i,B_j]_+ \rangle + 
\dlangle 
[A_i,H],B_j \drangle.
\end{eqnarray}
Since there are no many-body terms in the Hamiltonian, the Eq.~(\ref{EOM}) 
 allow us to obtain  numerically  exact results including all the contacts (see 
details in Ref.~\cite{Vernek2}).   Once we have the local GFs, we can 
compute the local density of states of the QDs sites,
\begin{eqnarray}\label{Atomic:dos}
\rho_{i}^s(\varepsilon)=-\frac{1}{\pi}{\rm Im \ }[\dlangle 
d_{i,s};d^{\dagger}_{i,s}\drangle], 
\end{eqnarray}
that can be readily extracted from the matrix GF (\ref{GFmatrix}).

As far as the interdot coupling concerns, it is well known that 
 our double QD system exhibits two distinct regimes: the atomic regime, 
for $V_{12}\ll\Gamma$, and the molecular regime, for $V_{12}\gg\Gamma$. In the 
atomic regime, because the electrons is more bound to the individual QD 
orbitals, the spectral properties of the system is better understood in terms 
of local density of states given by the Eq.~(\ref{Atomic:dos}). On the other 
hand, in the molecular regime (in which electrons are shared between the QDs) 
the  spectral properties can be better understood in terms of molecular density 
of states. To analyze the later regime we define two molecular orbitals,
\begin{eqnarray}
d_{\pm,s}=\frac{1}{\sqrt{2}}(d_{1,s}\pm d_{2,s})
\end{eqnarray}
that provide us with the molecular density of states
\begin{eqnarray}\label{Mol:dos}
\rho_{\pm}^s(\varepsilon)=-\frac{1}{\pi}{\rm Im \ }[\dlangle 
d_{\pm,s};d^\dagger_{\pm,s}\drangle].
\end{eqnarray}
This expression can be written in terms of the local and non-local quantum dot 
GFs as
\begin{eqnarray}\label{Mol:dos1}
\rho_{\pm}^s(\varepsilon)\!&=&\!\frac{\rho_1^s+\rho_2^s}{2}\mp\frac
{1} {2\pi} {\rm Im }\left[\dlangle 
d_{1,s};d^\dagger_{2,s}\drangle\nonumber \right.\\
&& \qquad  \qquad  \qquad \qquad \left.+\dlangle 
d_{2,s};d^\dagger_{1,s}\drangle\right],
\end{eqnarray}
where the non-local GFs, $\dlangle 
d_{1,s};d^\dagger_{2,s}\drangle$ and $\dlangle 
d_{2,s};d^\dagger_{1,s}\drangle$  appearing on the rhs of this expression are 
easily obtained by our procedure. Note that the interdot GF vanishes as 
$V_{12}\rightarrow 0$. Therefore, in this 
case, the molecular density of states is just the average of the QD 
(``atomic'') 
density of states. 
 
\subsection{Majorana Green's function}\label{Majorana_GF}
From the theoretical point of view, it is useful to define the 
Majorana GF 
\begin{eqnarray}
 M_{\alpha i,\alpha i}^{s}(\varepsilon) \equiv \dlangle  \gamma_{\alpha 
i}^{s}; \gamma_{\alpha i}^{s} \drangle \quad \mbox{(for 
$\alpha=A,B$),}
\end{eqnarray}
where 
%
\begin{eqnarray}
\gamma^s_{Ai}=\frac{1}{2}\left(f_{i,s} + 
f_{i,s}^{\dag}\right) \label{majoranaOp1}\\
\gamma^s_{Bi}=-\frac{i}{2}\left(f_{i,s}-f_{i,s}^{\dag}
\right) \label{majoranaOp2}
\end{eqnarray}
%
are the Majorana operators with the property 
\begin{eqnarray}
\gamma^s_{\alpha i}= \left(\gamma_{\alpha i}^{s}\right)^\dag 
\end{eqnarray}
and  obeying the anti-commutation relation 
\begin{eqnarray}
\left[\gamma^s_{\alpha 
i},\gamma^{s^\prime}_{\alpha^\prime 
j}\right]_+\!=\!2\delta_{i,j}\delta_{\alpha \alpha^\prime}\delta_{s 
s^\prime}.
\end{eqnarray}

Once we have obtained the electron GF,  the definitions of the Majorana 
operators (\ref{majoranaOp1}) and (\ref{majoranaOp2}) allows us to write the 
retarded Majorana GF in terms of the regular and the retarded Gorkov's 
GF as
 \begin{eqnarray}
M_{A i,A i}^{s}(\varepsilon)
\!&=&\!\frac{1}{4}\left[\dlangle f_{i,s };f_{i,s }^{\dag} \drangle 
\!+\!\dlangle f_{i,s }^{\dag};f_{i,s} \drangle \! + \! \dlangle f_{i,s 
};f_{i,s } \drangle \right. \nonumber\\ 
&& \left.+\dlangle  f_{i,s }^{\dag};f_{i,s }^{\dag} \drangle \right] ,
 \end{eqnarray}
and
 \begin{eqnarray}
 M_{B i,B i}^{s}(\varepsilon)
\!&=&\!\frac{1}{4}\left[
\dlangle f_{i,s };f_{i,s }^{\dag} \drangle \!+\!\dlangle f_{i,s 
}^{\dag};f_{i,s } \drangle \! - \!\dlangle f_{i,s };f_{i,s } 
\drangle  \right. \nonumber \\
&&\left.-\dlangle f_{i,s }^{\dag};f_{i,s }^{\dag} 
\drangle\right] .
 \end{eqnarray}
We can now compute the Majorana local spectral function (MLSF) 
as
\begin{eqnarray}
 \mathcal{D}_{\alpha i}^{s}(\varepsilon)=-\frac{1}{\pi}{\rm Im}[M_{\alpha 
i,\alpha 
i }^{s }] .
 \end{eqnarray}
Here, $f_{i,s}$ represents any regular fermion in the 
system. In particular, for the quantum dots, $f_{i,s}$ 
stands for $d_{i,s}$. Although the MLSF cannot be directly accessed  in 
experiments, it will help us to tell whether the electron density of states 
are composed by two Majorana modes or by a single one.  

\section{Numerical results}\label{Numerical}
To obtain our numerical results we will set the hopping $t_1=t_2=t=10\ 
\meV$ and $\Gamma=5.0\times 10^{-5} t$. We also set $\ve_1=\ve_2=0$. Following 
previous studies with realistic parameters\cite{Mourik,Rainis}, we 
set $\Delta_1=\Delta_2=0.025t$, $t_{\rm SO}=0.07 t$. 

\subsection{Single dot configuration}

\begin{figure}[b]
\centering
\subfigure{\includegraphics[clip,width=3.4in]{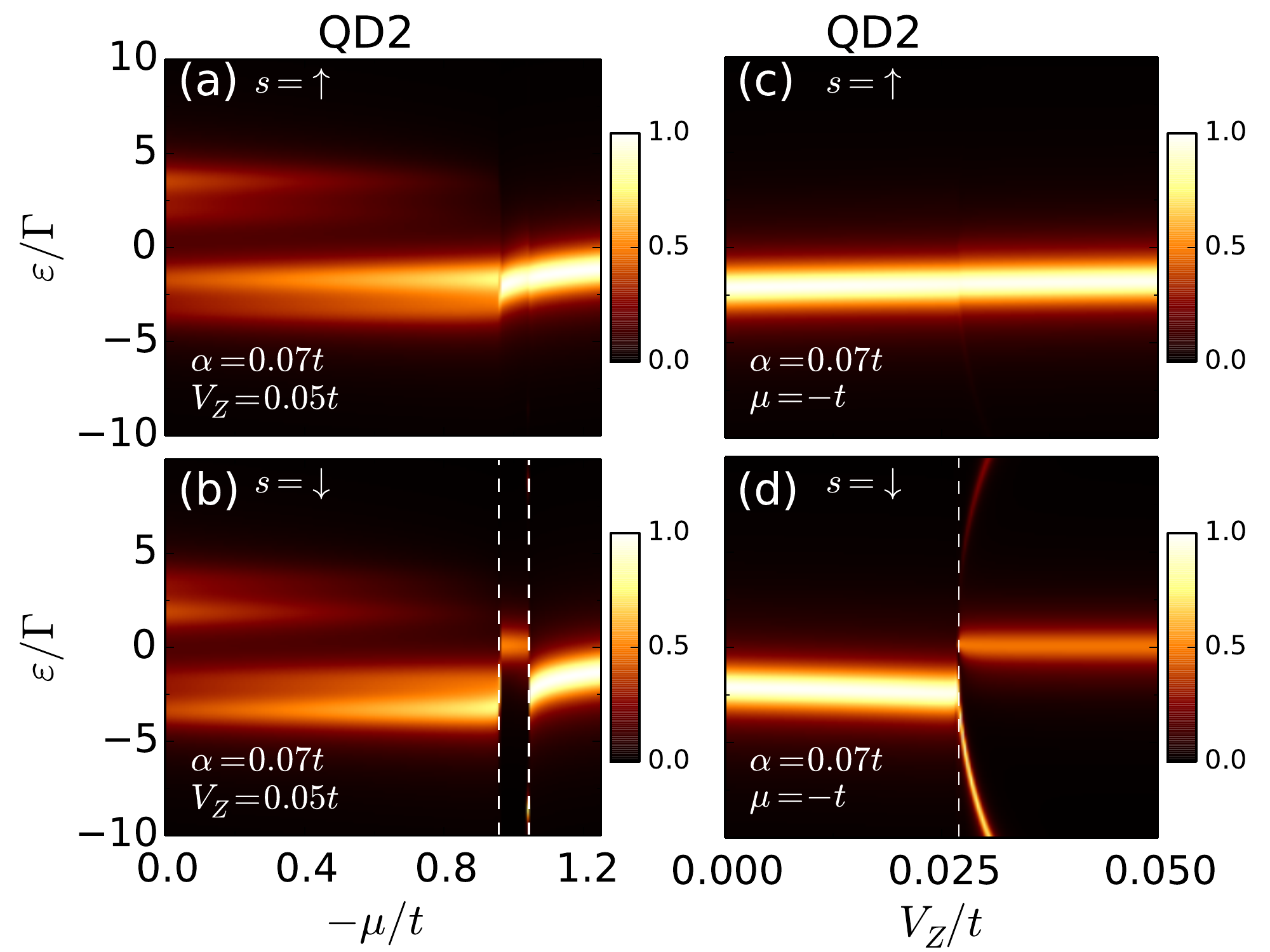}}
\caption{(Color online) Color map of the local density of states 
(in unity of $1/\pi\Gamma$) of the QD2 as a function of the energy $\e$ 
 and $\mu$ (left) and $V_Z$ (right). On the left, we have fixed $V_Z=0.05t$ 
and vary $\mu$ while on the right we fixed $\mu=-1.01t$ and vary $V_Z$. Top and 
bottom panels show the results for spin \emph{up} and \emph{down}, 
respectively. For all panels set $V_{12}=0$, so there is no influence of the 
QD1, and $V_1=V_2=0.008t$. The vertical dashed lines in the panels (b) and (c) 
delimit the boundary of the topological/trivial phase of the superconductor.} 
\label{single_dot}
\end{figure}

We first want to show that in the case of a single QD coupled to a 
superconductor, the regime in which there is a MBS in the system, the ABS are 
suppressed. To this end, let us consider in our system the situation in 
which the two QDs are decoupled from each other, by setting $V_{12}=0$.  Here 
we will show the suppression of the ABS in the QD2 as the SC2 is driven 
from its trivial to its topological phase. We do it by setting  
$V_z=0.05t$ and varying $\mu$. The results are shown in Fig.~\ref{single_dot}(a) 
and \ref{single_dot}(b), where we show the color map of density of states at the 
QD2 vs $\ve$ and $\mu$. Here, in particular, we set $V_2=0.008t$ so that the 
Andreev levels becomes very separated  and, therefore, more visible. For 
$\mu=0$ we clearly see the ABS symmetrically placed about $\e=0$ for both spin 
species. The small splitting in the ABS is due to the Zeeman effect.  Now, 
observe that as $\mu$ decreases from $0$ to $-t$ the upper branch of the ABS is 
progressively suppressed (also for both spins). This is because the strong 
electron-hole asymmetry induced by the change of the chemical potential 
suppresses the probability of creating a hole in the SC2. Interestingly, we 
note that, after the upper branch of the ABS has faded out, within a small 
range of $\mu$ (of the order of $V_Z$), we see the appearance of a peak at 
$\ve=0$ for spin \emph{down} only [Fig.~\ref{single_dot}(b)] (note that this 
feature is not seen for spin \emph{up}). This region is  precisely the 
topological phase of the SC2, in which the MBS leaks out from the SC2 into the 
QD2 (see discussion below). For further decrease of $\mu$, the SC2 returns to 
its trivial phase, in which there is no MBS nor ABS. The peaks of 
Figs.~\ref{single_dot}(a) and \ref{single_dot}(b) would be interchanged if we 
had chosen the opposite sign of $V_Z$. 

The small range of $\mu$ for which the SC2 is found in its topological 
phase can be readily understood from the topological condition $V_Z^2\geq \tilde 
\mu^2+\Delta^2$, or $\mu\in [-t-(V_Z^2-\Delta^2)^{1/2},-t+(V_Z^2- 
\Delta^2)^{1/2}]$. 
For $V_Z=0.05 t$ and $\Delta=0.025t$ we expect the MBS to appear only for $\mu$ 
approximately within the interval $[-1.043t,-0.957t]$ (or, equivalently, 
$-\mu/t$ within $[0.957,1.043]$) as seen in Fig.~\ref{single_dot}(b). Note that 
the zero mode is limited, precisely, within the two vertical dashed lines 
delimiting the calculated interval. The discontinuities are 
consistent with the  topological phase transition already predicted in 
Ref.~\cite{Alicea}. 

Now that we have shown the suppression of the ABS by the chemical 
potential, let us show how the ``surviving'' peak  evolves with $V_Z$, towards 
the MBS. In Figs.~\ref{single_dot}(c) and \ref{single_dot}(d) (for spin 
\emph{up} and \emph{down}, respectively) we show the density of states for the 
QD2 vs $\e$ and $V_Z$ for a fixed value of $\mu=-1.01t$, which is inside the 
region delimited by the two vertical lines in Fig.~\ref{single_dot}(b). Note 
that, essentially, nothing changes for the spin \emph{up} density of states. 
When $V_Z$ surpasses $V_Z^c$ there is a transition from the single to the 
tree-peak structure, in which the zero-energy peak corresponds to the MBS. The 
satellite peaks separate very quickly because the level of the QD2 is strongly 
coupled in resonance with the MBS from the wire. These peaks 
correspond to the Majorana state that is split off from the one that remains 
at energy zero. We will discuss this in the next subsection, when it will be 
more  visible with a smaller values of $V_{2}$. These results show that, 
indeed, ABS and the MBS cannot coexist in a single quantum dot. 

\begin{figure}[h]
\centering
\subfigure{\includegraphics[clip,width=3.3in]{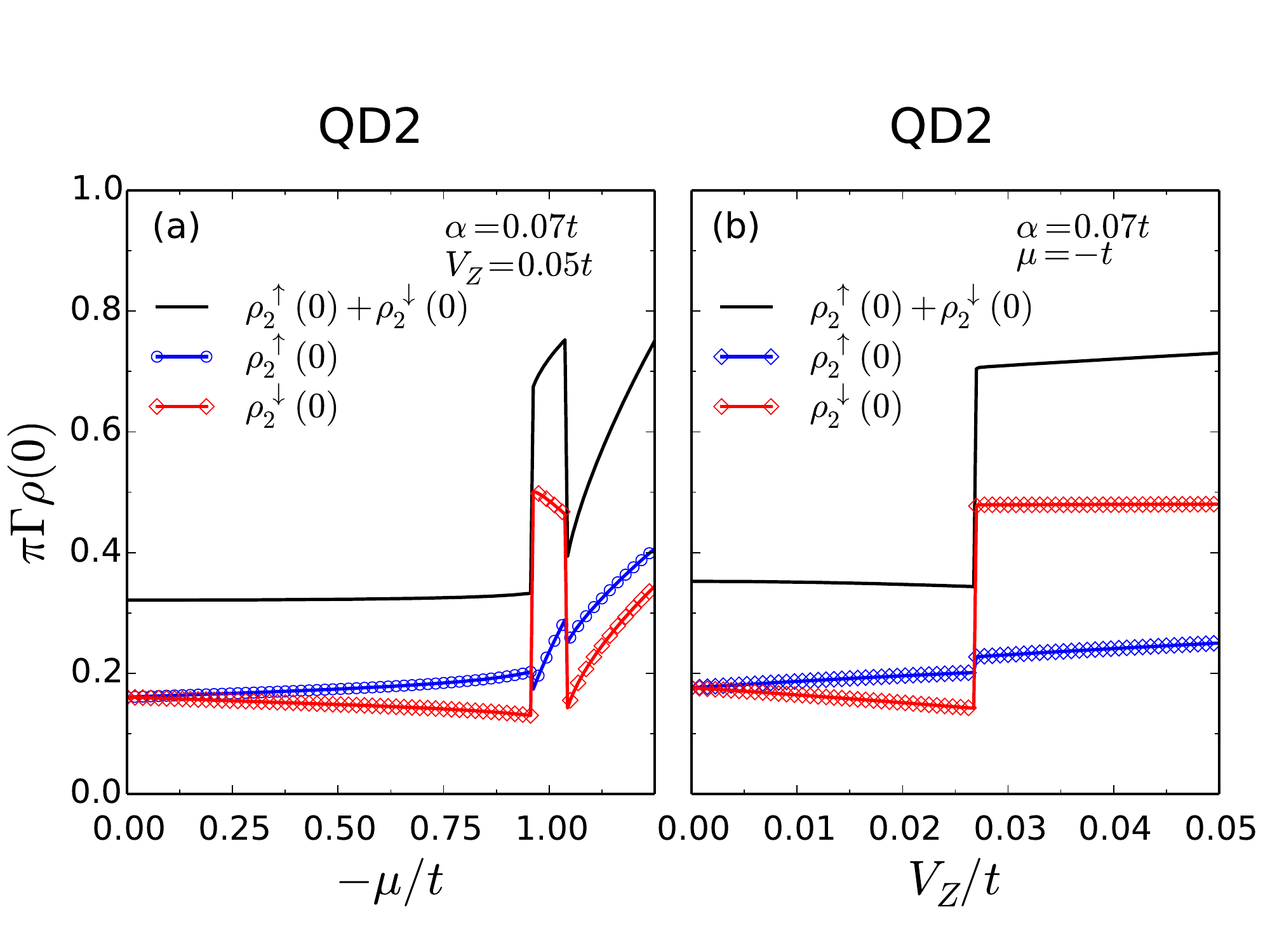}}
\caption{(Color online) Zero-energy LDOS vs chemical 
potential $\mu$ (left) and Zeeman energy $V_Z$ (right). The circles (blue) and 
the diamonds (red) curves correspond to the spin up and down, respectively, 
while the solid (black) lines corresponds to the total [$\rho_2^\up(0) 
+\rho_2^\dn(0)$] density of states. These curves correspond to a cut along $\ve 
= 0$ in the panels of Fig.~\ref{single_dot}. All the other parameters are the 
same as 
those of Fig.~\ref{single_dot}.} \label{tunn2}
\end{figure} 

As we have already mentioned above, experimentally, the signature of 
these features can be seen in the zero-bias differential conductance 
in spectroscopy tunneling measurements, that provide a probe of the LDOS at 
zero energy. To accomplish this, one can place an STM  tip on top of the QD2 to 
measure the zero-bias conductance from the normal contact into the lead 2 
into the tip. The observed conductance $G\propto 
\rho_2(0)=\rho^\up_2(0)+\rho^\dn_2(0)$ would be similar to what is shown in 
Fig.~\ref{tunn2} solid (black) curves. Setting the level of the QD2 $\ve_2$ at 
the Fermi level of the lead 2 and $V_Z$ finite, by varying $\mu$ one would 
observe a very low conductance in the ABS regime. When the SC2 enters the 
topological phase, the conductance jumps to a very higher value, as seen in 
Fig.~\ref{tunn2}(a). Then after leaving the topological phase, the conductance 
would drop suddenly to a smaller value. For completeness, in Fig.~\ref{tunn2}(b) 
we show the same quantity as a function of $V_Z$. The topological phase 
transition in this case would again be seen in a discontinuity in the 
conductance. In this case, there is only one jump because the SC2 is in its 
topological phase for all $V_Z>V_Z^{c}$. This plot is quite different from 
that one found in the Figs.~2(c) and 2(d) of Ref.~\cite{Liu}. Note that, in 
contrast with the result of Ref.~\cite{Liu}, here the conductance increases in 
the topological phase, while there the  conductance decreases.  This 
disagreement is because our approach accounts for the ABS fully, while in theirs 
there is no ABS.  

\subsection{Double dot configuration}
Since the Majorana and the Andreev bound states live in the system 
in different regimes of the single dot structure, it is still 
interesting to investigate  possible interplay between these bound states  with 
 MBS in the double QD configuration. Here, the SC1 is treated as a 
conventional superconductor (i.e., no spin-orbit, nor Zeeman field), whereas the 
SC2 is considered either in its normal phase ($V_Z<V_Z^{c}$) and in its 
topological phase  ($V_Z>V_Z^{c}$), with $V_Z^{c}=\sqrt{\tilde\mu^2+ \Delta^2}$. 
Then we make the coupling $V_{12}$ finite and study the SC2 in trivial and in 
the topological phase. Hereafter we set $\tilde\mu=-0.01t$ so that $V_Z^c\approx 
0.027$ and  use $V_Z=0.02t<V_Z^c$ for the trivial phase and $V_Z=0.05t>V_Z^c$ 
for the topological phase of the SC2. As long as $V_Z$ is larger or smaller 
than $V_Z^c$, the results we will show are quite independent of $V_Z$.

As we have seen above, the separation between the satellite peaks in 
the QD2 for the topological phase of the SC2 is much larger than the Andreev 
splitting, therefore, to make the results more visible, we now set the 
coupling between the dots and the superconductors $V_1=0.008 t$ and 
$V_2=0.002t$\cite{note1}.
For these parameters and  $V_{12}=0$ we have the following situation: the QD1 
will aways exhibit Andreev bound states because it is coupled to a normal 
superconductor. On the other hand, the QD2 will show a single particle peak if 
the SC2 is in its trivial phase ($V_Z<V_Z^{c}$) and a zero mode corresponding to 
the Majorana mode leaked from the SC2 in addition to a two split 
peaks, for ($V_Z>V_Z^{c}$). This three-peaks structure was extensively 
studied in our previous studies\cite{Vernek1,Vernek2}. 

\begin{figure}[t]
\centering
\subfigure{\includegraphics[clip,width=3.4in]{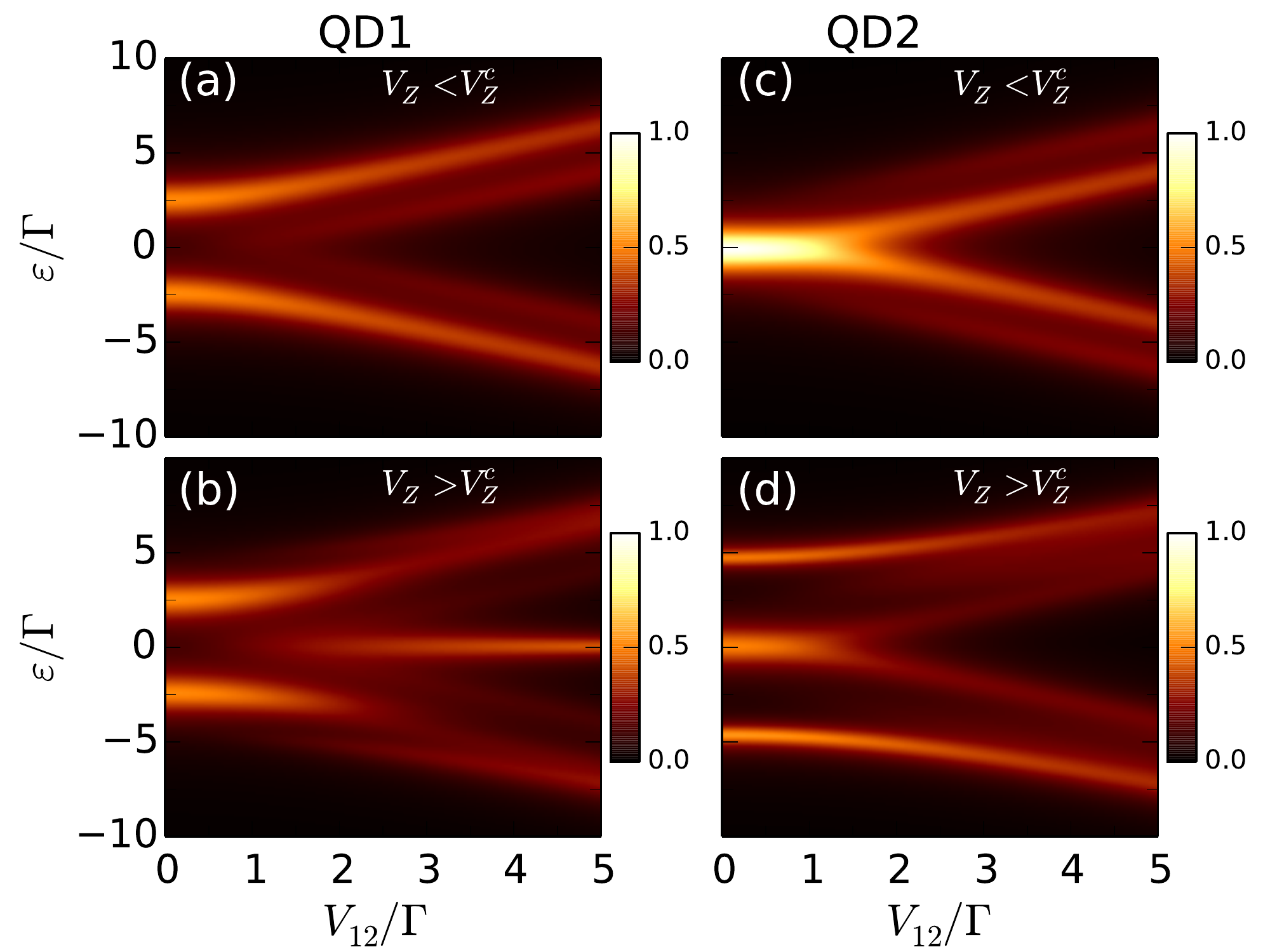}}
\caption{(Color online) Color map of the spin \emph{down} local density of 
states (in unity of $1/\pi\Gamma$) of the QD1 (left) and QD2 (right) 
as a function of the energy $\e$  and the interdot coupling $V_{12}$. Top and 
lower panels show the results for the trivial and topological phase of the SC2, 
respectively.} 
\label{dos1}
\end{figure} 

In Fig.~\ref{dos1} we show a color map of the local density of states of the 
QDs as a function of the energy $\ve$ (vertical axis) and the inter-dot 
coupling $V_{12}$ (horizontal axis). Let us start by analyzing 
these results in the normal phase of the superconductor SC2 ($V_Z<V_Z^c$, top 
panels). In this regime, $\rho^\dn_2$ [Fig.~\ref{dos1}(c)] has just one peak at 
$\ve=0$ while $\rho^\dn_1$ Fig.~\ref{dos1}(a)] has two peaks located 
symmetrically about $\ve=0$. These two peaks correspond to the Andreev bound 
states  discussed earlier. The single particle peak at $\ve =0$ observed in 
$\rho_2^\dn$ shows that the QD2 is not affected by the superconductors. The 
absence of Andreev states in the QD2 for $V_{12}=0$ is because the 
particle-hole asymmetry in the superconductor SC2 suppresses Andreev 
scatterings. As  $V_{12}$ increases, this scenario remains roughly unchanged in 
the atomic regime ($V_{12}\lesssim \Gamma$) and is progressively changed as the 
system enters its molecular regime $V_{12} \gtrsim \Gamma$.  In the molecular  
regime we note that $\rho_1^\dn$ exhibits a four-peak structure very much  
similar to those of $\rho_2^\dn$. These peaks correspond to the Andreev bound 
states that are split by the interdot coupling. Note that while the inner peaks 
are weaker in $\rho_1^\dn$ they are stronger in $\rho_2^\dn$, as compared to 
the outer ones. This is because the inner (outer) peaks of $\rho_1^\dn$ 
($\rho_2^\dn$) corresponds to the ABS in QD2 (QD1) projected onto QD1 (QD2). The 
peaks  appearing for positive and negative energies correspond to Andreev states 
on the ``$+$" and ``$-$" molecular orbitals, defined in Sec.~\ref{Electron-GF}. 
We will come back to this point and show the molecular density of states latter 
on in this section. 

\begin{figure}[b]
\centering
\subfigure{\includegraphics[clip,width=3.4in]{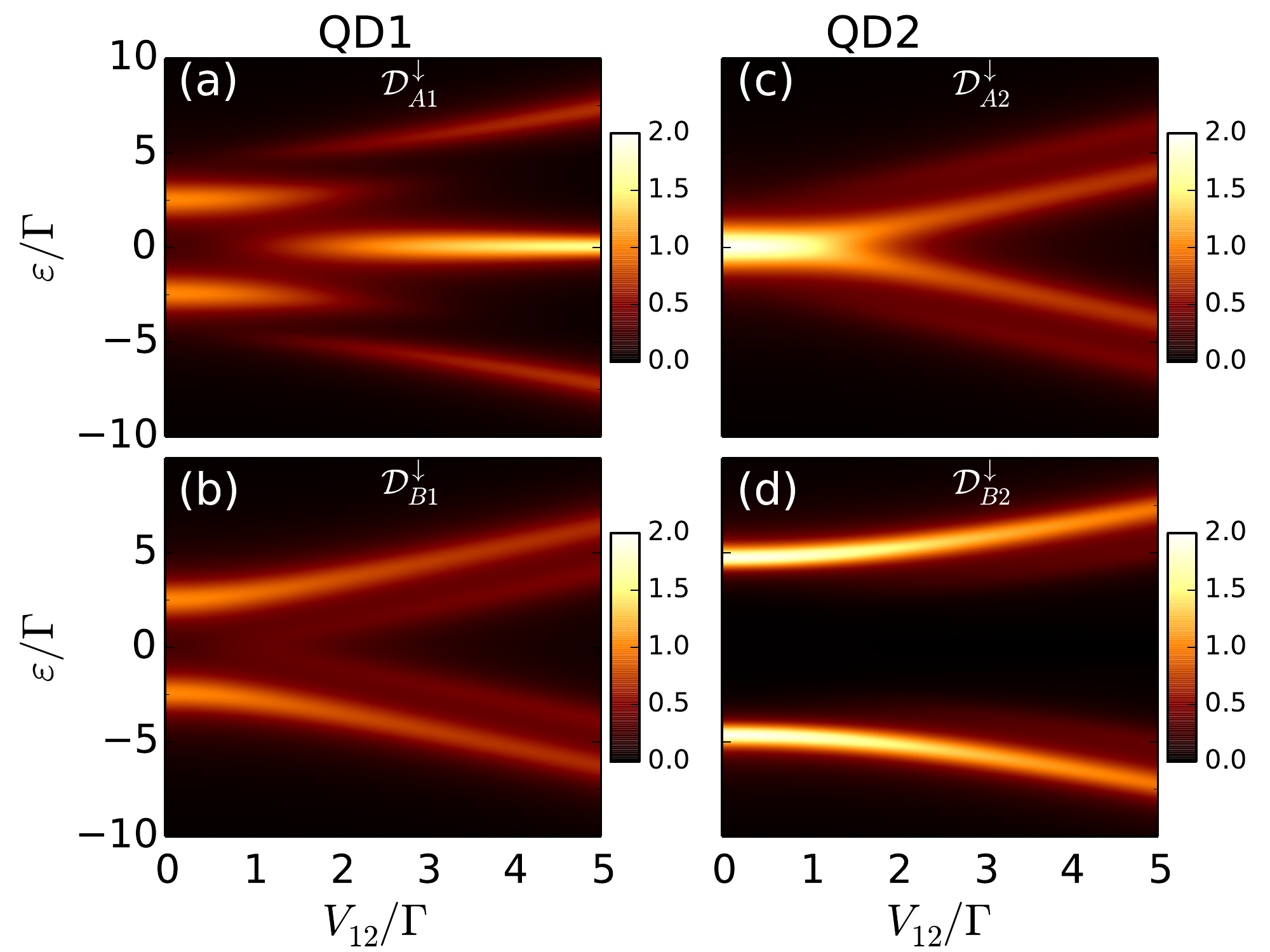}}
\caption{(Color online) Color map of the spin \emph{down} Majorana spectral 
function (in unity of $1/\pi\Gamma$) at the QD1 (left) and QD2 (right) 
vs the energy $\e$  and the interdot coupling $V_{12}$. Top and lower panels 
show the results for the topological phase of the wire 2, when we have a MBS. 
By comparing ${\cal D}_A^\dn$ with ${\cal D}_B^\dn$ for the same QD at the 
same energy, a peak appearing in only one of those shows there is a single 
Majorana mode. See discussion in the text.} 
\label{majdown}
\end{figure}

The results for the topological regime of the superconductor SC2 ($V_Z>V_Z^c$) 
are shown in Figs.~\ref{dos1}(b) and \ref{dos1}(d). We note in 
Fig.~\ref{dos1}(b) that $\rho^\dn_1$ is almost unchanged from the previous case 
in the atomic regime, with a two Andreev peaks in $\rho_1^\dn$. However, a 
complete different structure is seen in $\rho^\dn_2$. Note that even for 
$V_{12}=0$ we see the three-peaks structure known to be an indication of the 
leaking of the Majorana zero mode from the topological wire into the 
QD2\cite{Vernek1}. This scenario remains also roughly unchanged for the entire 
atomic regime ($V_{12}<\Gamma$).  Note that, because here the value of 
$V_2$ is smaller than in Fig.~\ref{single_dot},  here the satellite peaks are 
closer to each other as compared to those of Fig.~\ref{single_dot}(d) for 
$V_Z>V_Z^c$. Moreover, the satellite peaks observed in $\rho^\dn_2$ correspond 
to a Majorana state that is not bound to the QD2. Theoretically we can make it 
clear when we look at the Majorana spectral function as shown in 
Figs.~\ref{majdown} (shown only for $V_Z>V_Z^c$). Observe  that they appear only 
in ${\cal D}^\dn_{B2}$ [Fig~\ref{majdown}(d)] while the zero-energy peak is seen 
only for ${\cal D}^\dn_{A2}$ [Fig~\ref{majdown}(c)]. This shows that the nature 
of these peaks are completely distinct from the ABS. We should also emphasize 
that the zero-energy peak seen in Figs.~\ref{dos1}(c) corresponds to a regular 
fermion state whereas the on in Fig.~\ref{dos1}(d) corresponds to a Majorana 
bound state. Experimentally, the distinction between then can be made just by 
measuring the zero-bias conductance through the QD2 while changing the energy 
level of the QD2 in the regime of $V_{12}\rightarrow 0$. One would see that the 
conductance due to the Majorana peak would remain constant while in the regular 
fermion case it would show a peak. In fact, this corresponds a remarkable 
signature of the Majorana bound state as one of us have discussed in 
Ref.~\cite{Vernek1}.

The situation in the molecular regime ($V_{12}>\Gamma$) is the opposite. In this 
regime,  we see a three-peaks structure in $\rho_{1}^\dn$ and two ones in 
$\rho_{2}^\dn$. Note that as $V_{12}$ surpasses $\Gamma$, the Majorana zero 
energy peak dies off in the QD2 and arises in QD1 [compare the zero-energy peaks 
in the Figs.~\ref{dos1}(b) and \ref{dos1}(d)]. This can be attributed to a 
second leaking of the Majorana from the QD2 into QD1.  To support this 
interpretation, we now use the theoretical advantage of defining the Majorana 
GF. In Fig~\ref{majdown}(a) and Fig~\ref{majdown}(b) we show the spectral 
function of the Majorana operators $\gamma^\dn_{A1}$ (${\cal D}^\dn_{A1}$) and 
$\gamma^\dn_{B1}$ (${\cal D}^\dn_{B1}$), respectively, for the QD1 and in 
Figs~\ref{majdown}(c) and ~\ref{majdown}(d), the same for the QD2. By comparing 
${\cal D}^\dn_{A2}$ with  ${\cal D}^\dn_{B2}$ from Fig~\ref{majdown}(c) and  
Fig~\ref{majdown}(d), respectively, we see that for $V_{12}< \Gamma$ there is a 
zero-energy peak in ${\cal D}^\dn_{A2}$ but none in ${\cal D}^\dn_{B2}$.  On the 
other hand, there is no zero-energy peak in ${\cal D}^\dn_{A2}$ nor in ${\cal 
D}^\dn_{B2}$ for $V_{12}>\Gamma$. This indicates that there is  no single 
localized Majorana mode in the QD2. Now, if we look at QD1 Majorana spectral 
function [Figs.~\ref{majdown}(a) and \ref{majdown}(c)] we see that there is no 
Majorana zero mode for $V_{12}<\Gamma$. However, as  $V_{12}$ becomes larger 
that $\Gamma$, while the Majorana ${\cal D}^\dn_{A2}$ peak disappear from  the 
QD2 it emerges at the QD1, as shown in ${\cal D}^\dn_{A1}$. This is the second 
leaking we have mentioned above.      

\begin{figure}[b]
\centering
\subfigure{\includegraphics[clip,width=3.4in]{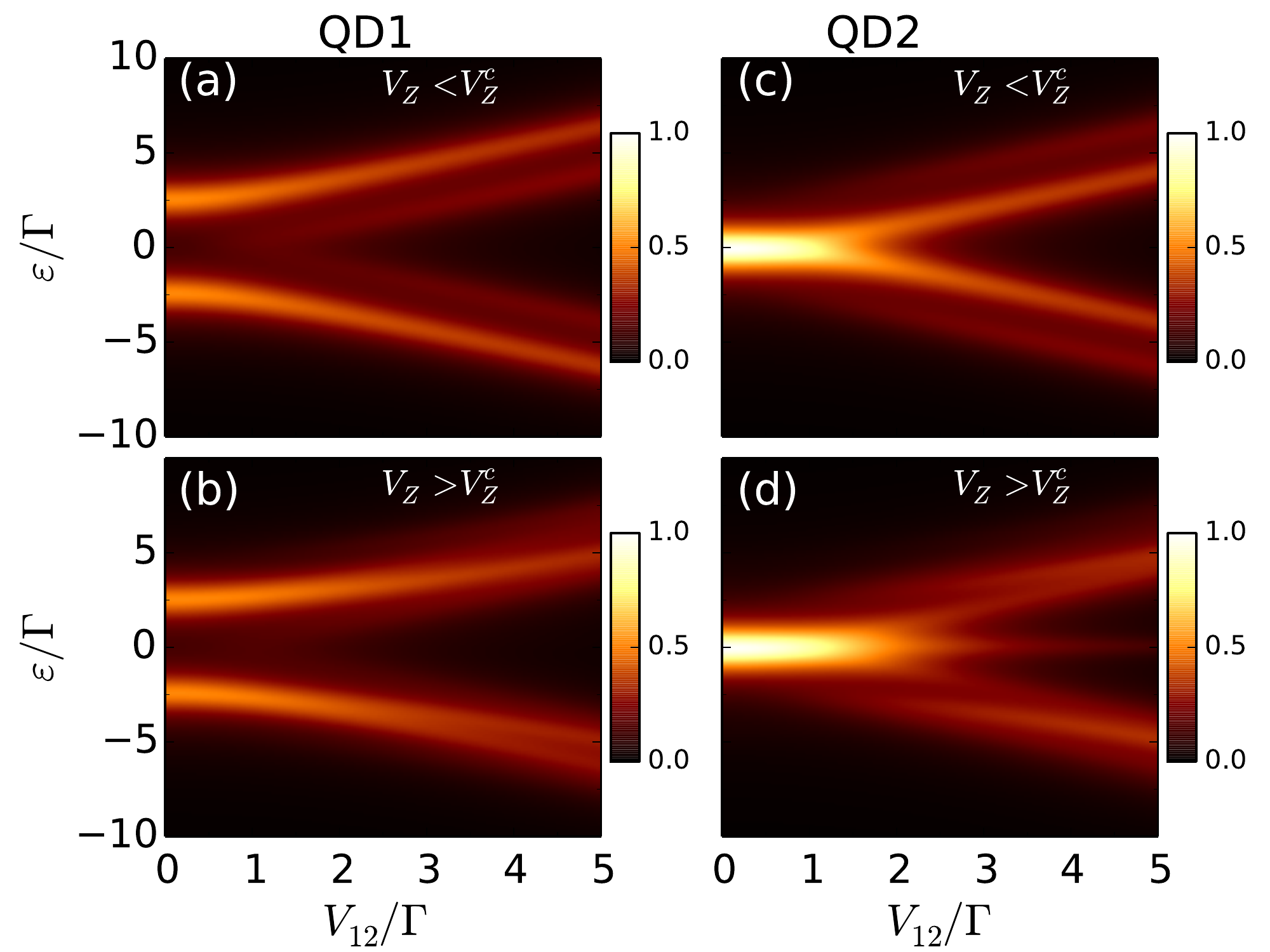}}
\caption{(Color online) Color map of the spin  \emph{up}  local density of 
states (in unity of $1/\pi\Gamma$) of the the QD1 (left) and QD2 
(right) as a function of the energy $\e$  and the interdot coupling $V_{12}$ for 
the same set of parameters as in 
Fig.~\ref{dos1}.} 
\label{dos1_up}
\end{figure} 

\begin{figure}[h]
\centering
\subfigure{\includegraphics[clip,width=3.4in]{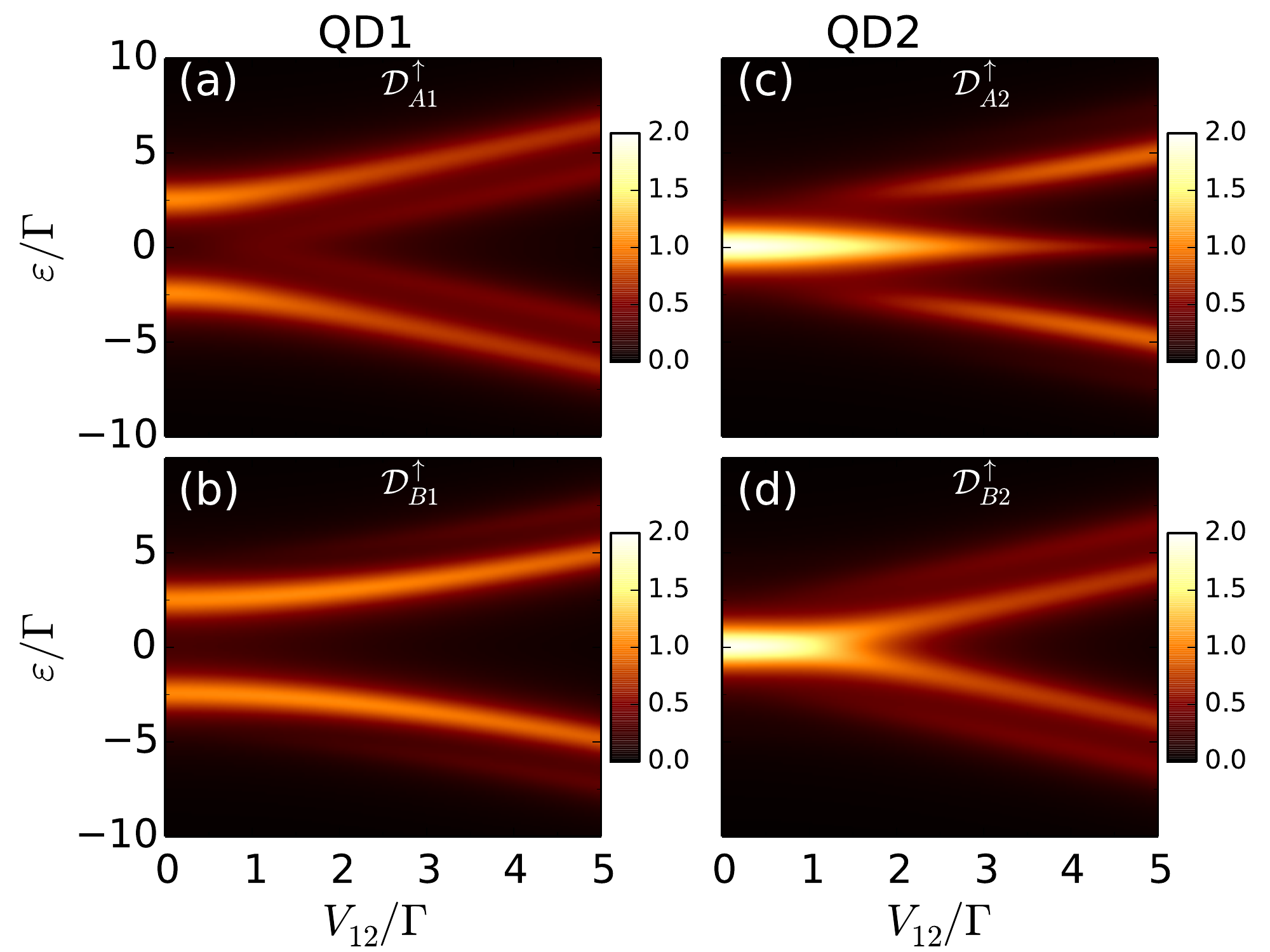}}
\caption{(Color online) Color map of the spin  \emph{up}  Majorana spectral 
function (in unity of $1/\pi\Gamma$) at the QD1 (left) and QD2 (right) 
vs the energy $\e$  and the interdot coupling $V_{12}$. Top and lower panels 
show the results for  topological phase of the SC2. Note the  ${\cal 
D}_A^\up$ , ${\cal D}_B^\up$ in each dot are very similar showing that there is 
no spin  \emph{up}  Majorana bound state in any of the QDs.} 
\label{majup}
\end{figure}

The reader may ask what happens to the density of states for the 
spin \emph{up}. To answer this question, in Fig~\ref{dos1_up} we show the spin 
{\it up} density of states in  for the same set of parameters as in 
Fig.~\ref{dos1}. Again, let us first look  at the trivial phase of the SC2 
($V_Z<V_Z^c$)  shown in Figs.~\ref{dos1_up}(a) and  \ref{dos1_up}(c) for the QD1 
and QD2, respectively. Comparing these plots with the corresponding ones in 
Fig.~\ref{dos1} we see that they are very much similar. This is because in 
the trivial  phase, the SC2 is overall gaped and the Zeeman effect has no 
effect on the rest of the system, as we have already pointed out earlier.  On 
the other hand, when we look at the the topological phase of the SC2, shown  in 
Figs. \ref{dos1_up}(b) and \ref{dos1_up}(d), and compare with 
Figs.~\ref{dos1_up}(a) and \ref{dos1_up}(c),  we see  some small effect for 
finite energy even in the spin  \emph{up}  electrons. This can be understood in 
the following way:  because we have chosen $V_z$ to be positive, the topological 
wire is strongly polarized with spin \emph{down}. Therefore, the 
coupling between electrons with spin  \emph{up}  in the dot and the Majorana in 
the SC2 is strongly suppressed. However, for finite $V_{12}$ the QD2 is 
indirectly coupled to the normal superconductor SC2 via QD1 and an electron with 
spin  \emph{up}  in the dot can couple to the Majorana mode via Andreev 
scatterings in the SC1. The effect is rather a small  since this corresponds to 
very high order processes.
This can also be seen in Fig.\ref{majup}, where we show the spin up
Majorana spectral functions. Note that the spin up Majorana
spectral functions for the QD2 shown in Fig \ref{majup}(c) exhibit a
signal of the zero mode coming from the spin down component.

\begin{figure}[]
\centering
\subfigure{\includegraphics[clip,width=3.4in]{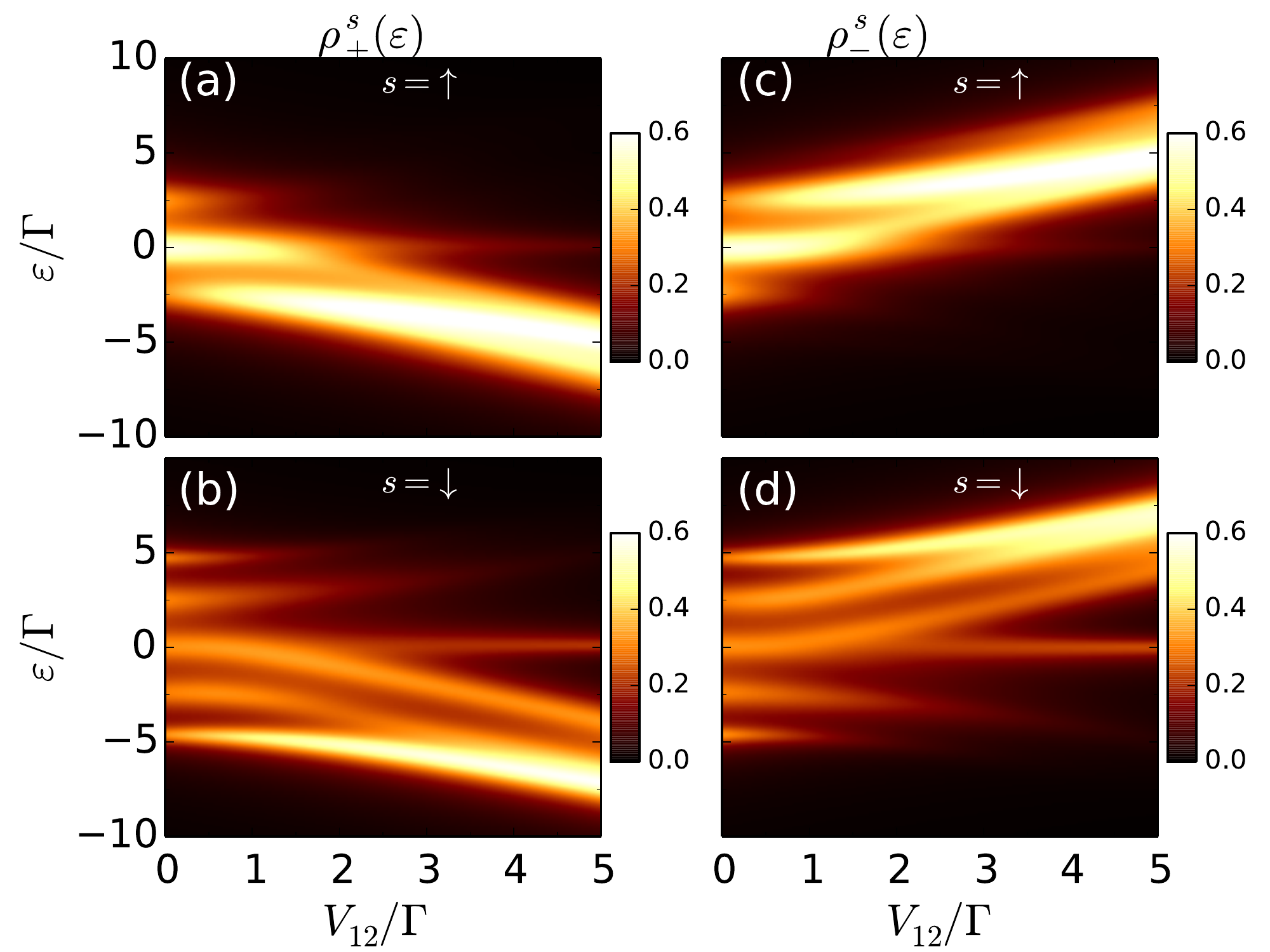}}
\caption{(Color online) Color map of the molecular density of 
states (in unity of $1/\pi\Gamma$) vs energy and the interdot 
coupling. Left and right panels show the density of states for the symmetric and 
antisymmetric orbital, respectively while top and bottom show for the spin 
$\up$ and $\downarrow$, respectively. One can see the MBS in both maps
characterized by the atomic regime. } 
\label{molec}
\end{figure}

To support our argument that in the molecular regime the Andreev states are 
bound to the molecule,  while the Majorana mode is bound to a particular QD 
(QD2, in our case),  let us show the density of states for the molecular 
orbitals [defined in Eq.~(\ref{Mol:dos1})].  In Fig.~\ref{molec} we show the 
molecular density of states for spin  \emph{up}  (top) and \emph{down} (bottom) 
and both molecular orbitals (left and right). Note first that for 
$V_{12}<\Gamma$ we see (for both orbital) three peaks for the spin  \emph{up}  
density of states and five peaks for the spin down. This is because in this 
regime, the molecular orbital corresponds, essentially, to the arithmetic mean 
value of the QD density of states shown previously. The central peak for 
spin \emph{down}, corresponding to the Majorana state bound to the QD2, now 
appears on both molecular orbitals.    
The molecular density of states are more useful for $V_{12}\gg\Gamma$. 
In this regime, we note that for spin  \emph{up}  [Fig.~\ref{molec}(a) and 
Fig.~\ref{molec}(c)]  there are  peaks only for negative energies in 
$\rho_+^\uparrow$ whereas only positive energy peaks appear in  
$\rho_-^\up$.  Similarly, $\rho_-^\downarrow$ and $\rho_+^\downarrow$ 
[Figs.~\ref{molec}(b) and \ref{molec}(d), respectively]  exhibit a similar upper 
and lower branch of peaks but with a slightly more complex structures, because 
of the splitting of the central peak due to the interdot coupling, but the 
positive and  negative energy  peaks  appear only in the ``$-$" and ``$+$" 
molecular orbitals, respectively. For $V_{12}\ll \Gamma$ their position in 
energy increases almost linearly with $V_{12}$, which is characteristic of 
molecular orbitals. Interestingly, we note that there is a zero-energy peak 
that remains in both molecular orbitals. This indicates that this state is 
always bound to a QD. Indeed, as we have seen previously, this zero-energy peak 
 corresponds to the Majorana mode bound to the  QD2. 

\begin{figure}[b]
\centering
\subfigure{\includegraphics[clip,width=3.4in]{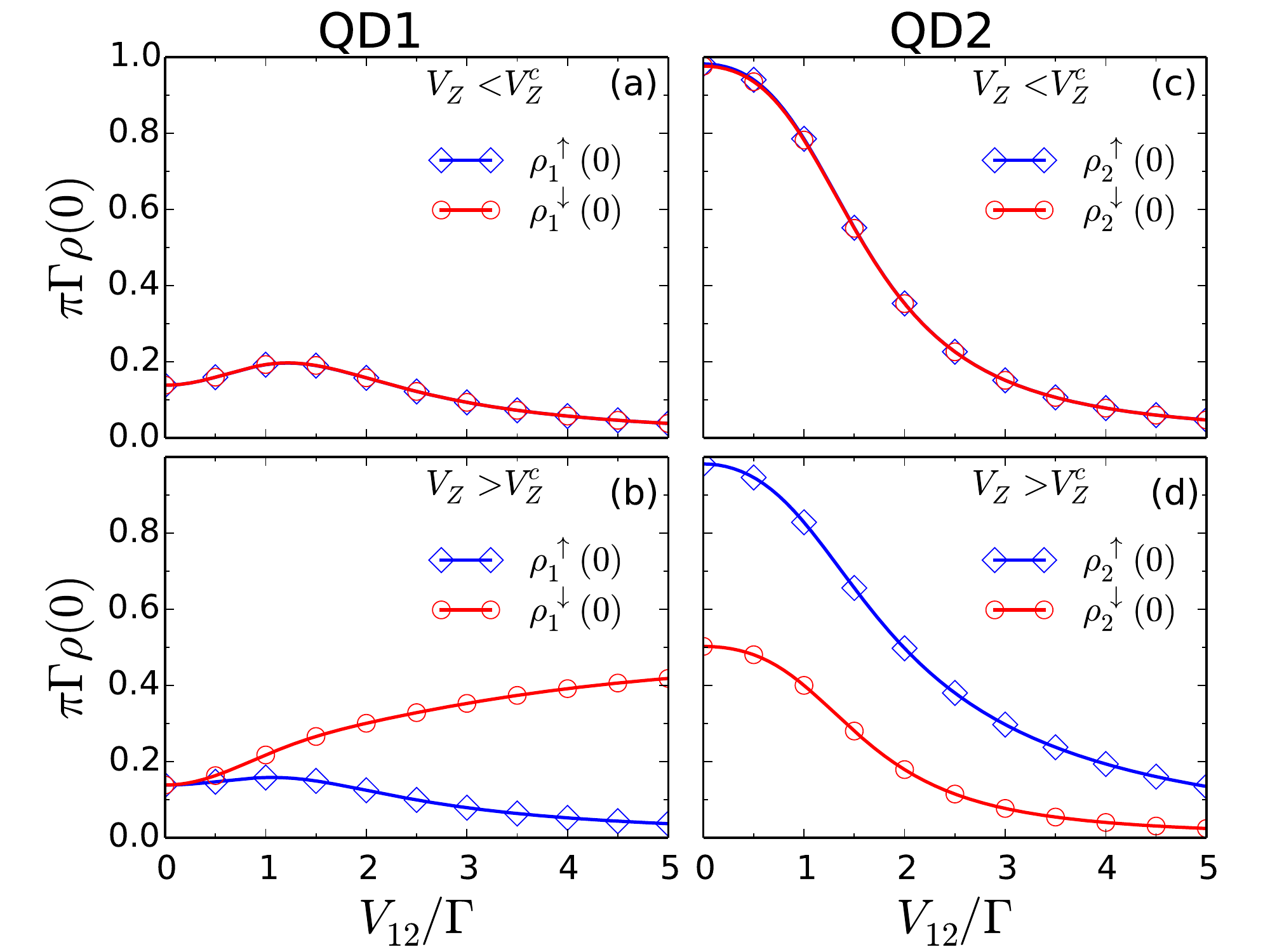}}
\caption{(Color online) Zero-energy LDOS of the QD1 (left) and QD2 (right) vs 
interdot coupling. Top and botton panels show the results for the trivial and 
topological phase of the SC2, respectively. All the other parameters are the 
same as in Fig.~\ref{dos1}.} 
\label{tunneling}
\end{figure} 

Similarly to what we have done in the single dot configuration, we 
now discuss how these features of the double-dot configuration could be observed 
within the current available experimental technology. This has to be done via 
local $dI/dV$ measurement as in Ref.~\cite{Goffman}. This is because in this 
configuration, the transport measurement between our lead 3  and 4 would not 
provide a local spectroscopy of the density of states of the QDs, individually. 
Therefore, similarly to the single-dot configuration, the STM tip can be placed 
on top of an individual QD. The zero-energy feature of our plots 
could be obtained experimentally by fixing the gate voltage of the dots and 
controlling the interdot coupling, while measuring the $dI/dV$ from the leads 
into the STM tip. Figure~\ref{tunneling} shows the zero-energy LDOS  vs 
interdot coupling for both trivial and topological phase of the SC2 for each QD, 
 individually. The features seen in this figure were already present in the 
previous one but here, some of them are more apparent and deserves to be 
discussed. First, as we already noted earlier, we see that in the trivial 
regime ($V_Z<V_{Z}^{c}$), even though the $V_Z$ is finite the DOS is degenerate 
for spin  \emph{up}  and \emph{down}. This is because the application of a 
small Zeeman field in the SC2 does not affect the LDOS of the dots since there 
is no in-gap states of the SC2 that couples to the dot. This is not the case in 
the topological phase of the SC2 in which there is a MBS. Further, note the 
difference between the LDOS of the QD1 and QD2. Note that for  $V_{12}\ll 
\Gamma$, $\rho_1^s(0)\ll \rho_2^s(0)$ (top panels). The suppression of the 
$\rho_1^s(0)$ for $V_{12}<\Gamma$ results from the presence of the Andreev bound 
states in the QD1. For $V_{12}\gg \Gamma$, we see that both $\rho_{1}^s(0)$ 
(both spins) decreases as the Andreev bound states take over the entire 
molecule. In the topological regime (bottom), the presence of the MBS leaked 
into the QD2 is manifested for $V_{12}=0$ as $\pi\Gamma\rho_1^\dn(0)=1/2$, in 
agreement to what some of us have reported in Ref.~\cite{Vernek1}. As $V_{12}$ 
increases, we see that $\rho_2^\dn(0)$ decreases while $\rho_1^\dn(0)$ 
increases, approaching $1/2$, because the MBS  leaks into the QD1. 
Finally, as remarkable signature of the MBS, when we look only into 
the QD1 we note that while $\rho_1^\up(0)$ decreases and $\rho_1^\dn(0)$ 
decreases for $V_{12}\gg \Gamma$, which produces strong polarization that could 
be detected using a spin polarized STM tip.

\begin{figure}[b]
\centering
\subfigure{\includegraphics[clip,width=3.4in]{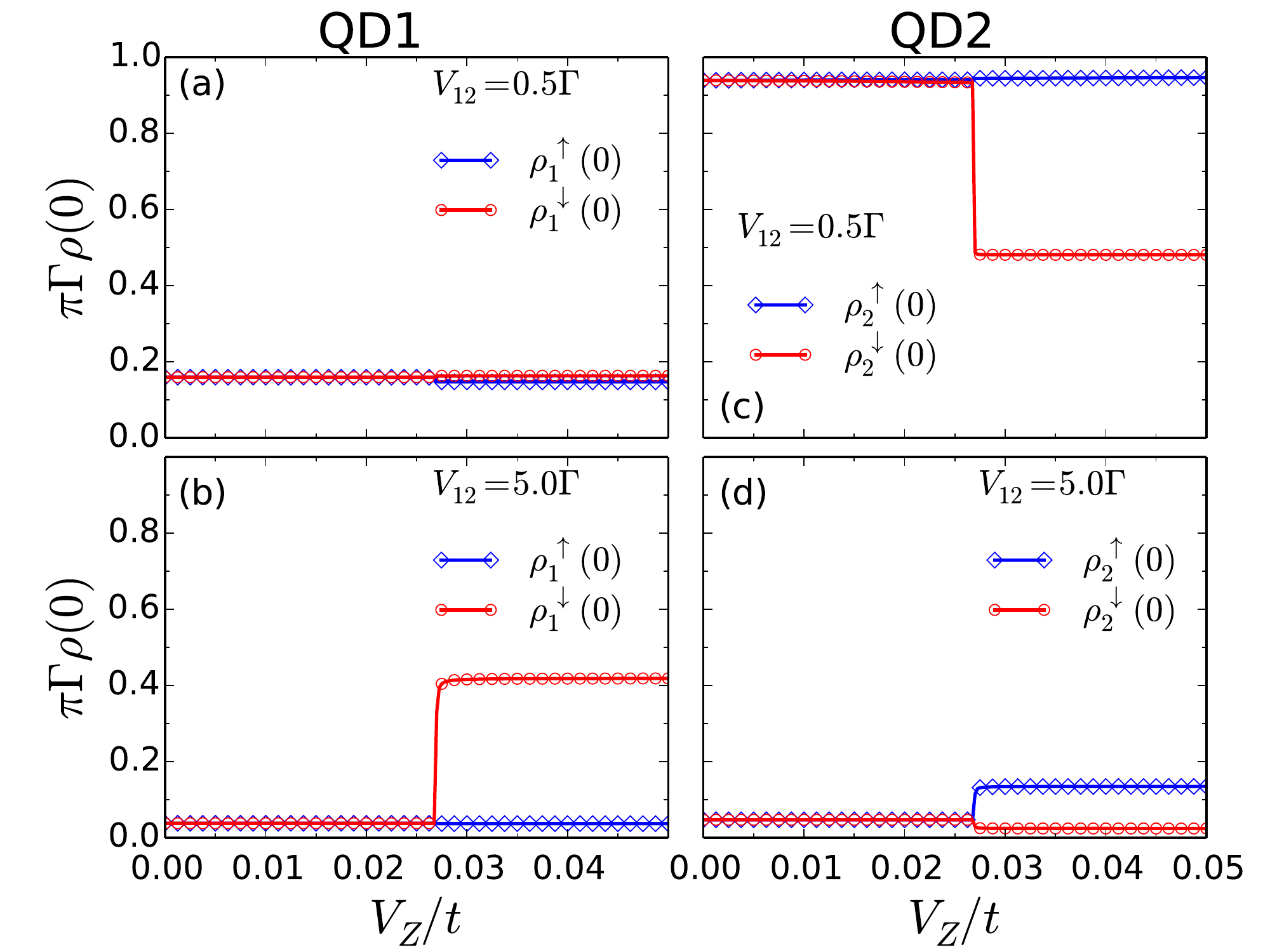}}
\caption{(Color online) Zero-energy LDOS of the QD1 (left) and QD2 (right) vs 
zeeman energy $V_z$. Top and botton panels show the results for the atomic 
regime ($V_{12}=\Gamma/2$) and  molecular regime ($V_{12}=5\Gamma$), 
respectively. All the other parameters are the same as in Fig.~\ref{dos1}.} 
\label{figure10}
\end{figure} 

Another interesting point is that  the signature of the 
topological quantum phase transition is still seen in the local density of 
states in the double dot configuration. To show this, in Fig.~\ref{figure10} we 
show the zero-energy density of states of the individual quantum dots vs $V_z$ 
for the both, the atomic ($V_{12}=\Gamma/2$) and the molecular regime 
($V_{12}=5\Gamma$) in the top and lower panel, respectively. In the atomic 
regime (top panels) we see that a jump in the $\rho_2^\dn(0)$ at 
$V_z=V_z^c$. On the other hand, in the molecular regime (lower panels) the jump 
is overwhelmingly more pronounced  in $\rho_1^\dn(0)$ [Fig~\ref{figure10}(b)] 
than the one observed in  $\rho_2^\dn(0)$ [Fig~\ref{figure10}(d)]. For the spin 
\emph{up} local density of states [diamonds (blue)], for both regimes and in 
both QDs we do not see any substantial change. The small jump in $\rho_2^\up(0)$ 
 noticed in Fig.~\ref{figure10}(d) [diamond (blue) line] results from a 
weak mixture of spin \emph{up} and \emph{down} in molecular regime due to 
Andreev scatterings. The results shown in Fig.~\ref{figure10} tell us that the 
topological quantum phase transition, well discussed in the single dot case, 
manifests itself even when the MBS and the ABSs coexist in the double double 
dot configuration. Moreover, alike in the single dot system, these features can 
be experimentally accessed in a simple spectroscopic tunneling measurement.

\section{Conclusion}\label{Conclusion}

Summarizing, we have studied the appearance of Andreev and Majorana bound states 
in a double quantum dot system. By studying the decoupled QDs configuration, we 
have shown that the MBS and the ABS cannot coexist in a single dot. This is 
because in the low-density regime (necessary for the topological phase 
of the superconductor) the electron-hole asymmetry  suppresses the ABS.  In the 
coupled dot configuration,  one of the QDs  (QD1) is coupled to a conventional 
superconductor that provides the Andreev bound states  while the other (QD2) is 
coupled to a topological superconductor, providing the Majorana bound state in 
its topological phase. We show that in the trivial phase of the SC2, in the 
atomic limit of the system ---where the interdot coupling is much smaller than 
the level broadening of the QDs ($V_{12}\ll \Gamma$)--- the QD1 exhibits Andreev 
bound states while the QD2 exhibits a regular fermion state broadened by 
$\Gamma$. When the system is brought into its molecular regime 
($V_{12}\gg\Gamma$), the Andreev bound states take over the entire molecule. 
More interestingly behavior occurs  in the topological phase of the SC2. In this 
situation, we show that for $V_{12}\ll\Gamma$ there are Andreev bound states in 
the QD1 for both spin components, while  in the QD2 we find a Majorana bound 
state  for spin \emph{down} and Andreev bound states for spin \emph{up}. In the 
molecular regime, on the other hand, the Majorana bound state leaks from the QD2 
into the  QD1, while the Andreev bound state can be seen in the entire molecule 
for the  spin  \emph{up}  component. These results reveal that, differently from 
the single dot case, the Majorana and the Andreev bound states coexist; while 
Andreev bound states can appear in molecular orbitals, the Majorana bound states 
is always bound to an atomic orbital. Another important feature that merits to 
be highlighted here is the remarkable signature of the topological quantum 
phase transition, manifested as a sharp jump in the zero-energy local density 
of states of the quantum dots, persisting  even in the double dot case, in 
which the Majorana and the Andreev bound states coexist. Finally, we believe 
our results are relevant from both theoretical and experimental viewpoints and 
should stimulate future spectral tunneling measurements to detect the signatures 
of  Majorana and Andreev bound states in the double dot system proposed here.

\ack
We would like to thank the Brazilians agencies CAPES, CNPq (Grant N. 
449488/2014-4) and FAPEMIG (Grant N. APQ-02344-14) for financial support. EV 
thanks C. Egues, L. G. G. V. Dias da Silva and D. Ruiz-Tijerina, for enriching 
comments and discussions.
\section*{References}


\begin{thebibliography}{<num>}
\bibitem{note1} Different values of $V_1$ and $V_2$ 
would essentially modify the positions and the intensity of the Andreev 
peaks.
\end{thebibliography}
\end{document}